\documentclass[preprints,article,accept,moreauthors,pdftex]{Definitions/mdpi} 

\firstpage{1} 
\pubvolume{1}
\issuenum{1}
\articlenumber{0}
\pubyear{2021}
\copyrightyear{2021}
\externaleditor{{Academic Editor}:} 
\datereceived{} 
\dateaccepted{} 
\datepublished{} 
\hreflink{https://doi.org/} 
\newcommand{\mJybeam}{mJy beam$^{-1}$}
\newcommand{\muJybeam}{$\mu$Jy beam$^{-1}$}
\newcommand{\HI}{H{\,\scriptsize I}}

\Title{Combining LOFAR and Apertif Data for Understanding the Life Cycle of Radio Galaxies }

\TitleCitation{Combining LOFAR and Apertif Data for Understanding the Life Cycle of Radio Galaxies}

\Author{Raffaella Morganti 
 $^{1,2,*}$\orcidA{},  Nika Jurlin $^{1,2}$\orcidH, Tom Oosterloo $^{1,2}$\orcidB, Marisa Brienza $^{3,4}$, Emanuela Orr\'u$^{1}$, Alexander  Kutkin $^{1,5}$\orcidC, Isabella Prandoni $^{4}$,  Elizabeth A. K.~Adams $^{1,2}$\orcidD, Helga  D\'enes $^{1}$\orcidE, Kelley M.~Hess $^{1,2}$\orcidF, Aleksandar Shulevski $^{6}$, Thijs ~van~der~Hulst $^{2}$\orcidG and Jacob Ziemke $^{1,7}$ }

\AuthorNames{Raffaella Morganti, Firstname Lastname and Firstname Lastname}

\AuthorCitation{Morganti, R.; Jurlin, N.; Oosterloo, T.A.;  Brienza, M.; Orr\'u, E.; Kutkin, A.; Prandoni, I.; Adams, E.A.K.; D\'enes, H.; Hess, K.M.; et al.}

\address{%
$^{1}$ \quad ASTRON, the Netherlands Institute for Radio Astronomy, Oude Hoogeveensedijk 4, 7991 PD Dwingeloo, The Netherlands;  morganti@astron.nl (R.M.); jurlin@astro.rug.nl (N.J.); oosterloo@astron.nl (T.O.); orru@astron.nl (E.O.); kutkin@astron.nl (A.K.), adams@astron.nl (A.K.A.), denes@astron.nl (H.D.), hess@astron.nl (K.M.H.), jacob.ziemke@icloud.com (J.Z.) \\
$^{2}$ \quad Kapteyn Astronomical Institute, University of Groningen, Postbus 800,
9700 AV Groningen, The Netherlands; j.m.van.der.hulst@rug.nl (J.M.vdH.) \\
$^{3}$ \quad Dipartimento di Fisica e Astronomia, Universit\`a di Bologna, Via P.\ Gobetti 93/2,     I-40129 Bologna, Italy; m.brienza@ira.inaf.it (M.B.) \\
$^{4}$ \quad INAF---Istituto di Radio Astronomia, Via P.\ Gobetti 101, I-40129 Bologna, Italy; prandoni@ira.inaf.it (I.P.) \\
$^{5}$ \quad Astro Space Center of Lebedev Physical Institute, Profsoyuznaya Str.\ 84/32, 117997 Moscow, Russia \\ 
$^{6}$ \quad Leiden Observatory, Leiden University, PO Box 9513, NL-2300 RA Leiden, The Netherlands; shulevski@strw.leidenuniv.nl (A.S.)\\
$^{7}$ \quad  University of Oslo Center for Information Technology, P.O. Box 1059, 0316 Oslo, Norway
}

\corres{Correspondence: morganti@astron.nl}

\abstract{
Active galactic nuclei (AGN) at the centres of galaxies can cycle between periods of activity and of quiescence. Characterising the duty-cycle of AGN is crucial for understanding their impact on the evolution of the host galaxy. For radio AGN, their evolutionary stage can be identified from a combination of morphological and spectral properties.
We summarise the results we have obtained in the last few years by studying radio galaxies in various crucial phases of their lives, such as remnant and restarted sources. We used morphological information derived from LOw Frequency ARray (LOFAR) images at 150 MHz, combined with resolved spectral indices maps, obtained using recently released images at 1400 MHz from the APERture Tile In Focus (Apertif) phased-array feed system installed on the Westerbork Synthesis Radio Telescope.
Our study, limited so far to the Lockman Hole region, has identified radio galaxies in the dying and restarted phases. We found large varieties in their properties, relevant for understanding their evolutionary stage. We started by quantifying their occurrences, the duration of the `on' (active) and `off’ (dying) phase, and we compared the results with models of the evolution of radio galaxies. In addition to these extreme phases, the resolved spectral index images can also reveal interesting secrets about the evolution of apparently normal radio galaxies. The spectral information can be connected with, and used to improve, the Fanaroff--Riley classification, and we present one example of this, illustrating what the combination of the LOFAR and Apertif  surveys now allow us to do routinely.
}

\begin{document}
\section{Introduction}
\label{sec:Introduction}

\textls[-15]{Since the first studies were conducted on active galactic nuclei (AGN), it has been suggested that the release of energy from  super-massive black holes (SMBH) goes in cycles,  with alternating periods of activity and periods of quiescence, see, e.g., \cite{Woltjer59,Marconi04,Saikia09,Morganti17} and references therein. In recent years, this characteristic of AGN has become even more relevant because of the role the energy released by the SMBH (in its active phase) can have on the gas content and star formation of the host galaxy. According to cosmological simulations, describing galaxy formation and evolution \cite{Bower06,Schaye15,Weinberger17,Dave19}, this impact of the AGN has to be recurrent, repeating multiple times in the life of a galaxy (also known as the {\emph{feedback cycle}}) in order to reproduce the growth of the SMBH, as well as the star formation history we observe in the Universe (se,e e.g., \cite{Ciotti10,Gaspari13,Gaspari17}).}

An active SMBH can manifest itself in different ways, depending on various properties of the host galaxy \cite{Kauffmann04,Best05}.  The presence of radio emission and radio jets is one of these manifestations. In radio galaxies, the presence of aged and recurrent activity is long known. This has been mostly supported by the serendipitous discovery of objects, such as  double--double radio galaxies, i.e., radio galaxies with pairs of radio lobes along the same projected jet axis, which are interpreted as the result of interrupted and restart jet activity,  \cite{Schoenmakers00,Saikia09,Konar13,Mahatma19},  or by studies selecting objects with specific properties, for example those having an ultra steep radio spectrum, i.e., where the electrons are not replenished because the nuclear activity has stopped and, therefore, considered to be dying sources (see, e.g., \cite{Parma99,Murgia11} and the discussion below).  


However, a more complete and quantitative characterisation of the {\emph{life cycle of radio galaxies}},  the cycling between active, dying and restarting phases, requires a more systematic search and larger samples, especially for catching the most elusive phases in this cycle: the remnant and restarted phases. This has only now become possible thanks to the \emph{ large and deep} radio surveys at low radio frequencies that are becoming available. 

To achieve the goals of obtaining larger (and less biased) samples of remnant and restarted radio galaxies, and a more systematic and quantitative assessment of the life cycle, two elements are important: (1) being able to trace the morphology and structures on different scales (i.e., the presence of extended, low surface brightness amorphous features, while at the same time,  having high spatial resolution to identify the presence of compact structures); and (2) obtaining the (resolved) spectral information down to the lowest frequencies. The relevance of tracing low surface-brightness structures was already highlighted in the study of \cite{Saripalli12} and is motivated by the fact that, when the activity stops, every compact structure (such as cores, hot spots, etc.) disappears fairly quickly {(as short as $\sim1$ Myr, \cite{Mahatma18})}, while more slowly, the larger radio galaxy lobes  expand and lose energy, becoming fainter. {Deriving the time scales of the latter is one of the goals of our project}.
At the same time, a relatively high spatial resolution allows us to properly separate these diffuse components from any compact structures, if present. Images of the  spectral index $\alpha$ (defined through $S \propto \nu^{-{\alpha}}$, where $S$ is flux density and $\nu$ the frequency) distribution across a source can act as a `fossil record' tracing the history of the radio emission over long timescales (see, e.g., \cite{Morganti17} for more details). 
The low-frequency emission  (i.e., below 1~GHz) is the last part of the spectrum that is affected by radiative losses, meaning it is detectable for a longer period, while it can also still contain the signature  of the injection spectrum of the original electrons.   
Based on  synchrotron theory, we know that, in a system of particles emitting synchrotron radiation under the effect of a magnetic field, the energy loss is higher for high-energy particles ($dE/dt \propto \nu^2$). This means that, over time,  the spectrum steepens, starting from the higher frequency end. Thus, the presence of a break in the power-law shape of the spectrum with a steepening of the  spectral index at higher frequencies  can been used as an indicator for the ageing of the electrons. Furthermore, if the steepening is particularly large, with spectral indices steeper than $\alpha = 1.2$ (i.e., ultra-steep spectrum, USS), it is the signature of plasma devoid of any fresh particle injection \cite{Komissarov94}. Therefore, this provides a key signature for tracing remnant radio galaxies: inactive or dying radio AGN where the central engine has stopped or its activity is substantially dimmed. 
The lower  the frequency at which this ultra steepening occurs, the larger the time-span (for a given magnetic field) since the last re-acceleration of the electrons has occurred and, thus, the older the remnant source is.

Finally, the co-existence of USS emission from `old' plasma with an active central region can be used to identify situations in which different populations of electrons co-exist originating from different phases of activity, as in restarted radio AGN (see below the case of 3C~388).
Although uncertainties related to, for example, non-uniform magnetic fields, can complicate the interpretation, observing extreme spectral indices can provide unambiguous evidence of both stopped and restarted activity. 

\section{LOFAR and Apertif: The New Possibilities they Offer}
\label{sec:surveys}

Based on what we have mentioned above, the morphology of  sources traced at low frequencies by deep surveys, combined with resolved spectral index studies down to low frequencies, are powerful tools for achieving the goal of characterising the life cycle of radio galaxies. We are doing this by searching for remnant and restarted radio galaxies that correspond to the most elusive and least known phases of the life cycle. 
For our study we mostly use the  LOw Frequency ARray (LOFAR, \cite{Haarlem13}) and the APERture Tile In Focus (Apertif) surveys \citep{Cappellen21}, although other radio telescopes, in particular the Giant Metrewave Radio Telescope (GMRT) and the Very Large Array (VLA), can be particularly useful for follow-up observations (e.g., at higher spatial resolution).

\subsection{LOFAR Surveys and Deep Fields}

LOFAR is, by now, a mature radio telescope that has provided a large number of interesting new results to understand radio sources (see, e.g., \cite{Hardcastle20}). For the present study, an important characteristic of LOFAR  is the fact that, next to the presence of long baselines giving high angular resolution allowing to identify compact structures, LOFAR has a dense coverage of short baselines making it sensitive to low surface brightness emission. 
Important for building good statistics of rare objects is the fact that
LOFAR is performing large surveys as well as deep observations of famous fields. 
The LOFAR Two-Metre Sky Survey (LoTSS) at 150 MHz aims to cover the entire northern sky and the first data release is described in \cite{Shimwell19}. 
The images produced have a median noise level of $S_{\rm 144\ MHz} = 71$ $\mu$Jy  beam$^{-1}$ and a spatial resolution of 6$^{\prime\prime}$.
The deep images of famous fields, among which the Lockman Hole (LH) field used for the project presented here, reach median sensitivity of $S_{\rm 144\ MHz} \le 30$ $\mu$Jy  beam$^{-1}$  (e.g., \cite{Tasse21}, see also the A\&A Special Issue {(Volume 648, 2021)} describing the data release of the images).

LOFAR can  observe at frequencies even lower than 150 MHz and the LOFAR LBA Sky Survey (LoLSS) aims to cover the entire northern sky in the frequency range  \mbox{42--66 MHz}, with a resolution of 15$^{\prime\prime}$. This survey is  ongoing and a first  data release, albeit at lower spatial resolution than the final survey (45$^{\prime\prime}$) due to limitations in the calibration and pipeline, as published in \cite{deGasperin21}.

\subsection{The Apertif Surveys at 1400 MHz}

The APERture Tile In Focus (Apertif) phased-array feed (PAF) system was recently installed at the Westerbork Synthesis Radio Telescope (WSRT) and operates at 1400~MHz (see \cite{Cappellen21}). Using the PAFs in the focus of each WSRT dish, 40 partially overlapping primary beams are formed in the sky simultaneously, which in practice means that  the system is equivalent to 40 old WSRT single receiver systems observing in parallel and, thus, having a much larger field of view. The total area covered in one observation by such a mosaic is about
$3.5^\circ \times 3^\circ$, reaching a spatial resolution of about $12^{\prime\prime} \times 12^{\prime\prime}/\sin\delta$. Apertif is used as a survey instrument, and the Apertif imaging surveys started on 1 July, 2019, with the aim of covering up to $\sim2500$ square degrees of the northern sky.

The first data from the Apertif surveys---providing radio continuum, \HI\, and polarisation images and cubes---have become publicly available in the first data-release, which was done in November 2020. More details are given in the documentation accompanying this release\endnote{http://hdl.handle.net/21.12136/B014022C-978B-40F6-96C6-1A3B1F4A3DB0}. The released continuum images produced by the current pipeline reach typical rms noise of $\sim 40$ $\mu$Jy (Adams et al.\ in prep, Kutkin et al.\ in prep).
Because of their resolution, observing frequency, and sensitivity, the Apertif surveys provide an ideal complement to the surveys done with LOFAR. 
In particular,  the  spatial resolution of the final images from the lower-frequency LOFAR survey, LoLSS \cite{deGasperin21}, will be fully comparable to the one of Apertif; thus, providing an ideal match, covering from 54 to 1400 MHz.

\section{Lessons Learned from Single Object, Low-Frequency Studies}
\label{sec:SingleObjects}

Before addressing the selection and study of larger samples, we briefly discuss some highlights of what we have learned in recent years from single-object studies and, in particular, from adding the low frequency information from LOFAR.
Such studies have allowed us to explore candidate remnant and restarted radio galaxies in order to expand our knowledge of their properties. 
Here, we briefly summarise the results obtained for three objects. These results  will then be used to guide the selection criteria for obtaining larger samples as described in Section~\ref{sec:LargeSamples}.

The first object is the so-called \emph{ blob 1}, presented in \cite{Brienza16}. This object was discovered with LOFAR as a large (700 kpc), amorphous, and low surface-brightness structure. The remnant nature was first suggested based on the morphology and then confirmed by the sharp frequency break observed at high frequencies (>1400 MHz). With multi-frequency observations, we traced the radio spectral energy distribution (SED) and found not only an extreme spectral index above 1400~MHz (typical of remnant sources), but also a very large spectral index curvature (SPC) between frequencies below and above 1400~MHz (see \cite{Brienza16}). The modelling of its spectrum with standard models of radiative evolution allowed us to constrain the age and the duration of the active vs. dying phases. 
 
In the case of blob 1, a relatively short active phase (15 Myr) was followed by a significantly longer time ($\sim 60$ Myr) spent in the remnant phase. This source is a remnant located in a low-density environment, i.e., outside a galaxy cluster, unlike where most of the remnant sources were considered to reside \cite{Murgia11}. Despite having confirmed (based on their spectrum) that its large scale lobes are not fuelled anymore with fresh particles, we detect in blob 1 a faint core \cite{Brienza16}, suggesting that either the black-hole emission never really stops and  continues at low levels, or the nucleus has restarted its activity very recently. 

The results on blob 1 show that remnant radio galaxies exist that are not (yet) characterised by ultra-steep spectra (USS) at low frequencies, i.e., the extreme spectral indices have not yet reached the low frequencies. 
This can be taken as an indication of their relatively young age (since the activity stopped) and/or of a particularly low magnetic field in these remnant sources .  
Thus, from blob 1, we have learned that using USS as selection criterion may result in an incomplete inventory of  remnant radio galaxies and  that the presence of a weak core cannot be excluded in remnant sources.  

\begin{figure}
\widefigure
\includegraphics[height=6.5cm]{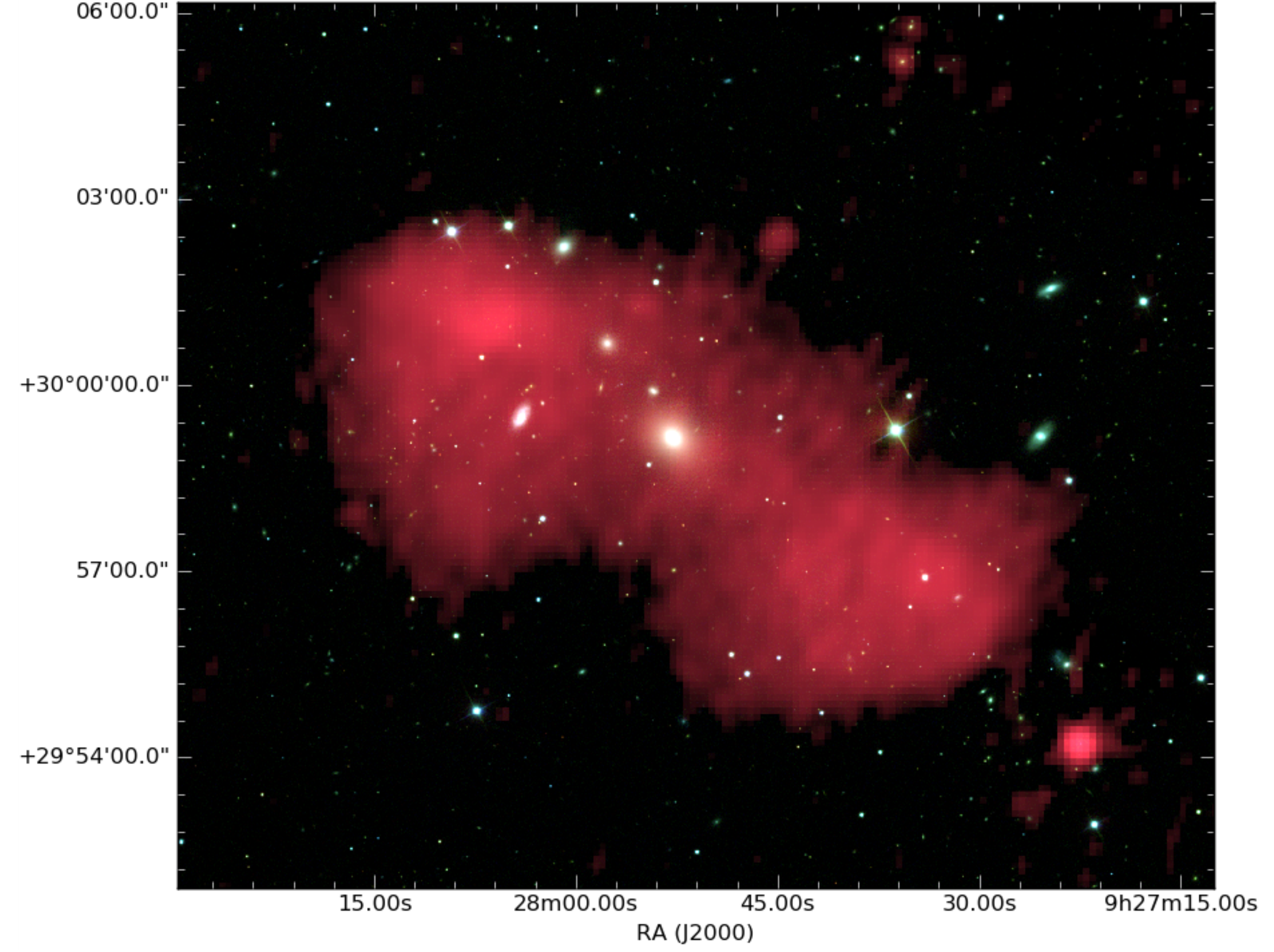}
\includegraphics[height=6.5 cm]{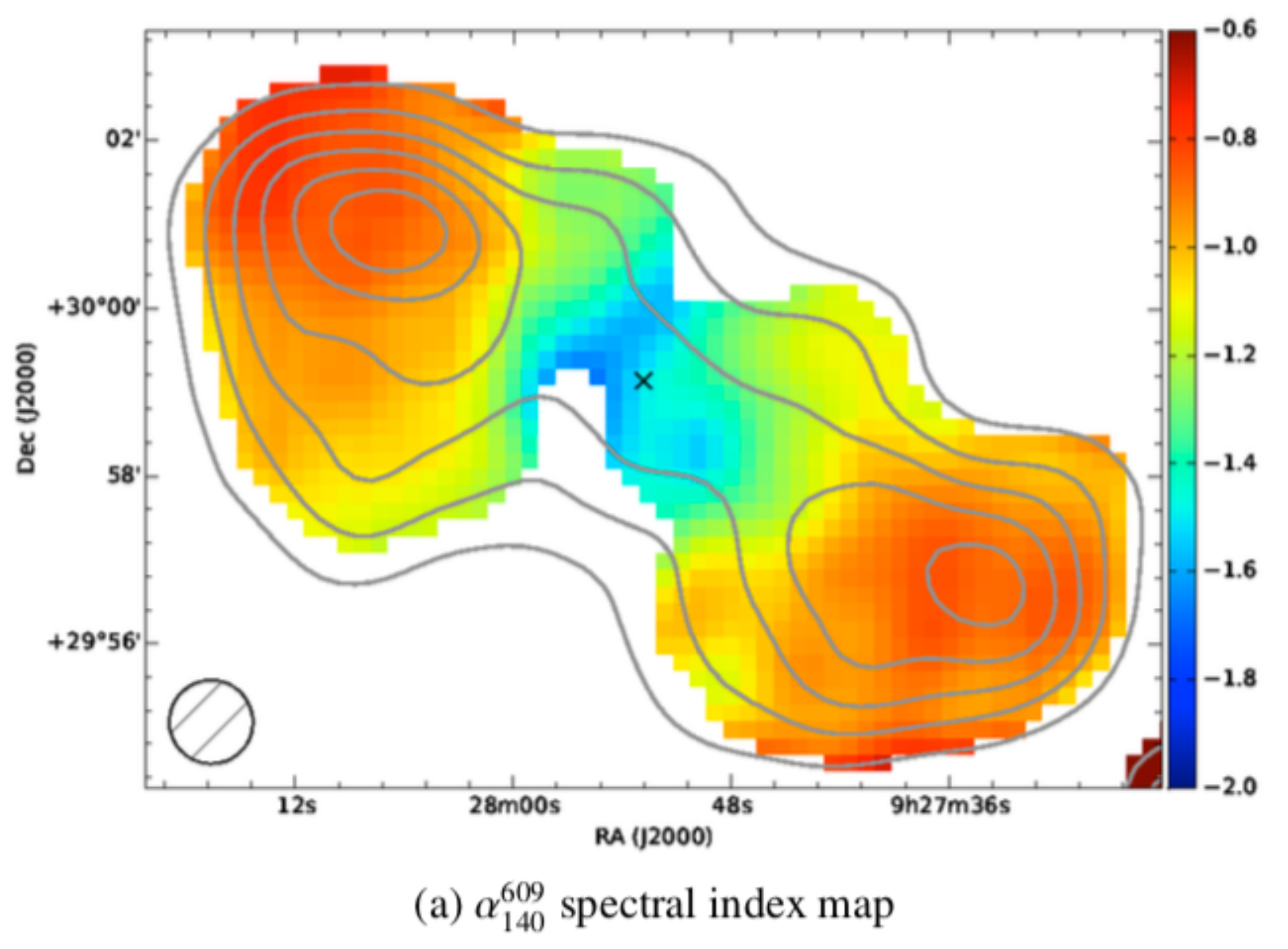}
\caption{Left: total intensity (red) of B2~0924+30 showing the diffuse, amorphous, and low-surface brightness structure superposed to an optical image (the galaxies are shown in white, background in black). Right: the spectral index between 140 and 609 MHz, showing that  flatter spectral indices are found towards the outer edges of the lobes, \cite{Shulevski17}. Figures taken from \cite{Morganti17,Shulevski17}.}
\label{fig:0924}
\end{figure}

The importance of adding the resolved spectral information down to low frequencies has also been demonstrated for the case of B2~0924+30, until recently and before the discovery of blob 1, the most well known remnant radio galaxy not residing in a cluster environment \cite{Cordey87}. 
The distribution of the total intensity and of the low-frequency spectral index from our LOFAR study \cite{Shulevski17} are shown in Figure \ref{fig:0924}.
They illustrate how the low-frequency spectral index can reveal more details on the nature and evolution of the source.  The addition of the low frequency information to that of the high frequencies studies \citep{Jamrozy04} has allowed us to study the spectral index properties over a large band (up to 5000~MHz). 
Using this combination, it was found that the lobe spectral index is somewhat flatter at low frequencies and, interestingly, the flatness is more pronounced at the outer edges of the lobes (see Figure \ref{fig:0924}).
The flatter spectral index at low frequency  and the spectral index gradient indicates a recent switch-off of the source. Furthermore, the flatter regions at the edges, reminiscent of old hot spots, suggests that B2~0924+30 could be a fading Fanaroff--Riley type II (FR-II) source, i.e., the youngest regions are found towards the outer edges of the lobes, and the oldest are the regions towards the host galaxy.
Thus, the spectral properties (more than the morphology) may give indication on the type of progenitor of the radio source and on the timescales of the evolution of the remnant sources. In the case of B2~0924+30, it suggests that, after $\sim$50 Myr, we can still see some difference in the spectral index at the location of the termination of the jets  \cite{Shulevski17}. 

The resolved spectral index images have also shown the possibility of tracing restarted activity. The radio galaxy 3C~388 was, until recently, the only object where two regions within the lobes with very different spectral properties were found \cite{Roettiger94}. This was taken as an indication of two epochs of activity---one representing remnant emission from a past phase of activity and one representing a recent restarted phase. This scenario was confirmed by our study expanding the frequency range using LOFAR data \cite{Brienza20} in which we derive a total source age of <80 Myr and a quite fast duty cycle of about $t_{\rm on}/t_{\rm tot} \sim 60$\%. The analysis also shows that mixing of electrons from the two phases may also be happening, complicating the age estimates. 

\end{paracol}

\begin{paracol}{2}
\switchcolumn


In summary, these studies have shown us that remnant and restarted radio galaxies can have a variety of properties and that a corresponding variety of selection criteria should be used in order to have a (as complete as possible) census of these objects.  They also illustrate that, although observations at more frequencies are necessary for a complete study, the low-frequency part is very  useful and provides important constraints.

  
\section{Building Large Samples of Remnant and Restarted Radio Galaxies } 
\label{sec:LargeSamples}

As discussed above, the new, deep radio surveys will allow us to build statistically interesting samples of remnant and restarted radio galaxies, and to derive, in a robust  way, their occurrence and their properties. 
A first project attempting this was carried out by~\cite{Jurlin20}, who selected sources in the Lockman Hole (LH) region over an area of about 20 deg$^2$---centred on RA 10$^{\rm h}$47$^{\rm m}$00$^{\rm s}$, Dec +58$^\circ$04$^\prime$59$^{\prime\prime}$. A description of the region and of the initial LOFAR work can be found in \cite{Mahony16}, while more recent and deeper LOFAR observations of this area are presented in \cite{Tasse21,Mandal21}.

\textls[-25]{The results described in Section\ \ref{sec:SingleObjects} provided the basis for applying a broader range of selection criteria to extract candidate remnant and restarted radio galaxies, which go beyond the selection of, respectively, USS extended sources and/or double--double radio galaxies.  }

For the selection of  remnant radio galaxies, in addition to USS properties, we consider  the lack of compact structures, the presence of extended low surface brightness emission at low frequencies, and a low core prominence (i.e., defined as CP = $S_{\rm core}/S_{\rm total}$). The above-mentioned criteria have proven to be successful. 
Indeed, for  most objects selected with these criteria, the classification as remnant radio galaxies has been confirmed for the majority of these objects  by  follow-up observations \cite{Jurlin21}. 

In addition, to identify candidate restarted radio galaxies, we combined the selections based on morphological properties (such as double--double structures) with cases where a prominent core (i.e., CP $> 0.1$; \cite{Jurlin20}) was embedded in extended low surface brightness emissions. This  follows the assumption that a remnant phase---characterised by the amorphous, low-surface brightness structure---is followed by a restarted phase that makes the central region particularly bright. We also selected, as candidate restarted sources, those that had steep spectrum cores (i.e., steeper than $\alpha=0.5$), which we took as an indication that the nuclear region hosts newly formed jets that were already quite prominent but that were not identified as such at the resolution of the available observations.  A comparison sample of active radio galaxies was also identified  in order to derive the relative occurrence of remnant and restarted radio galaxies and to compare the properties of the host galaxies among these various source classes. A full description of the selection criteria for the samples and of the results is given in \cite{Jurlin20} (see also Section\ \ref{sec:Results}). 

This work can now be complemented with the information on the resolved spectral index  obtained using the combination of LOFAR images at 150~MHz and Apertif images at 1400~MHz. 
A first proof-of-concept was presented in \cite{Morganti21}. Despite being limited to a relatively small area (about 6 deg$^2$ of the LH region), and despite using Apertif science verification data, this study did show the potential of adding the spectral information (even if limited to two frequencies only). 

Now that more Apertif observations are available, we can further investigate the properties of the resolved spectral index distribution for all of the remnant and restarted sources selected in our previous studies (\cite{Brienza17,Jurlin20} in the LH area covered by LOFAR).

\end{paracol}

\begin{figure}
\widefigure
\includegraphics[width=16.0 cm,angle=0]{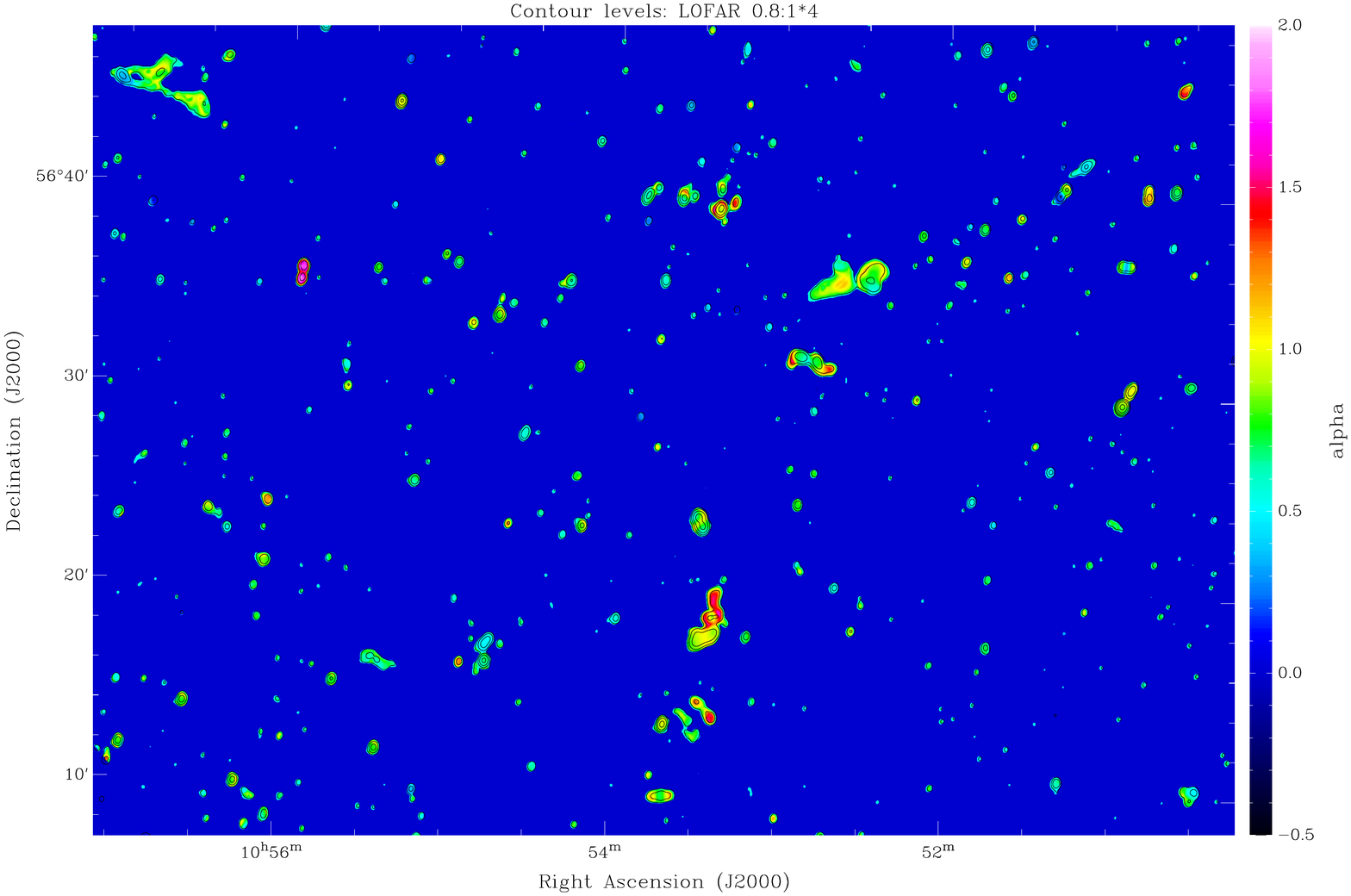}
\caption{A relatively small region (about 40 arcmin$^2$) extracted from the 20 sq deg  spectral index image of the Lockman Hole area. Spectral indices are derived between 150 and 1400~MHz (LOFAR-Apertif) with a spatial resolution  of \mbox{19 $\times$ 15 ArcSection}. The LOFAR image has been convolved to this resolution. The contours (from 0.8 \mJybeam\ to 1 Jy beam$^{-1}$ in steps of a factor 4) represent the LOFAR emission. The colour scale represents the spectral index values: red colours indicate  steep spectrum regions, blue colours the regions with flat spectrum. Some of the spectral indices are lower limits, i.e., in regions undetected by Apertif the limits are derived assuming for the 1400~MHz flux an upper limit of  5-$\sigma$, where $\sigma = 40$ $\mu$Jy. See text and Figure \ref{fig:NewExamples} for details.
Sources undetected by LOFAR, but detected by Apertif (i.e., with strong inverted spectrum), are not considered in this image.} 
\label{fig:RegionSI}
\end{figure}
\begin{figure}
\widefigure
\includegraphics[width=16.5 cm]{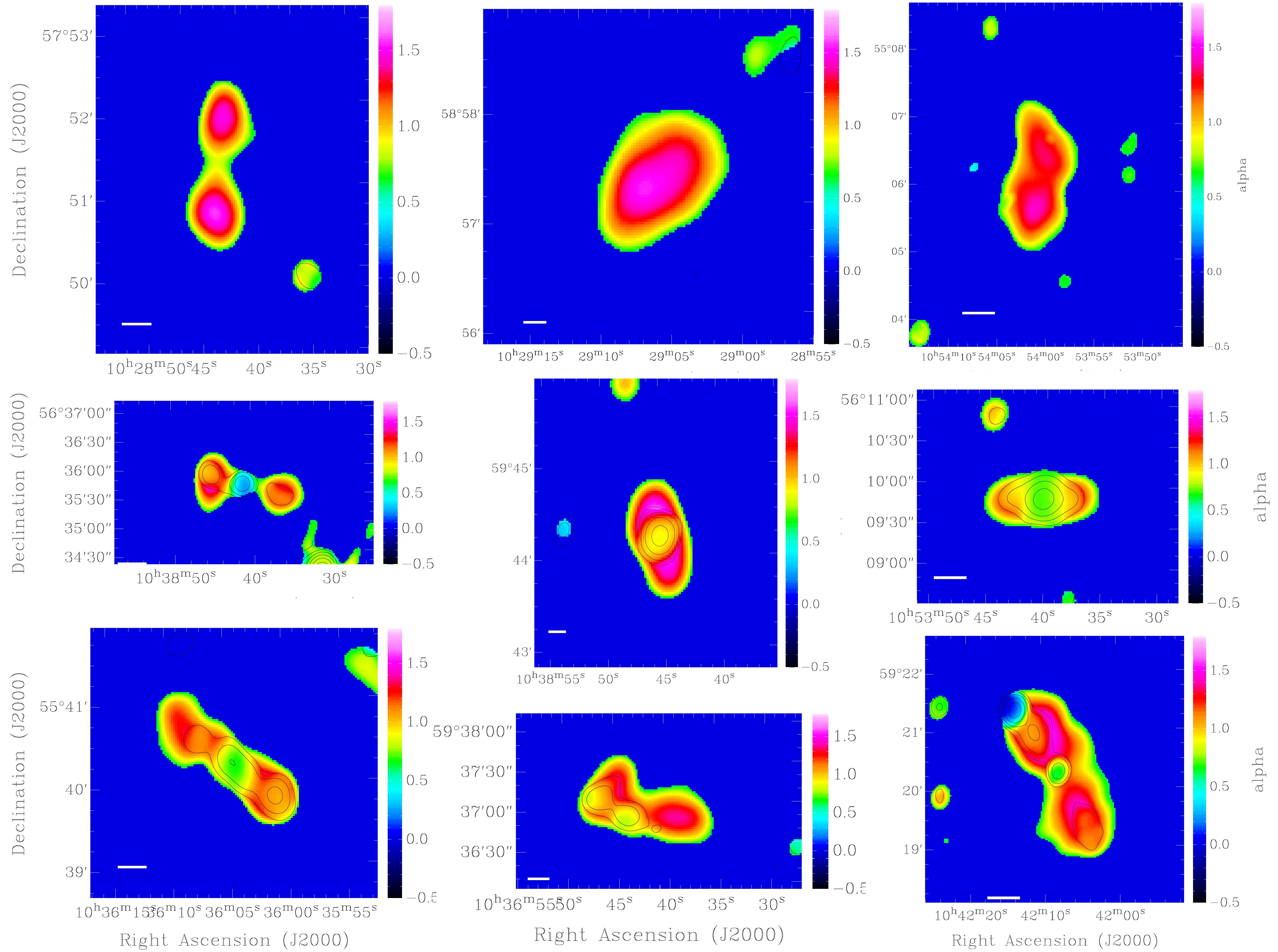}
\caption{Examples of resolved 150--1400 MHz spectral index images from the LH area, illustrating cases of remnant (top row), restarted candidates (middle row), and not-confirmed remnant radio galaxies (bottom row) from \cite{Brienza17,Jurlin20}, see text for details. The complex structure of the spectral index is clearly seen in some of these objects. The contours represent the 1400~MHz emission from Apertif. Regions where no contours are drawn, correspond to  Apertif non-detections. In these regions, the spectral index values are lower limits (i.e., the actual spectral index is steeper than what indicated).  For the three remnant sources in the top row, Apertif does not detect emission over the full source, and these are entirely USS sources.
The white bar indicates 100 kpc (assuming an average redshift of these sources, $z \sim 0.5$, see \cite{Jurlin20,Jurlin21}). The colour scale is the same as in Figure \ref{fig:RegionSI}.
The artificially flatter spectral index at the edge of the sources results from using a fixed value for the 1400~MHz flux  (5-$\sigma$) in case of Apertif non-detection, see Section\ \ref{sec:LH}.
}
\label{fig:NewExamples}
\end{figure}

\begin{figure}
\widefigure
\includegraphics[width=19 cm]{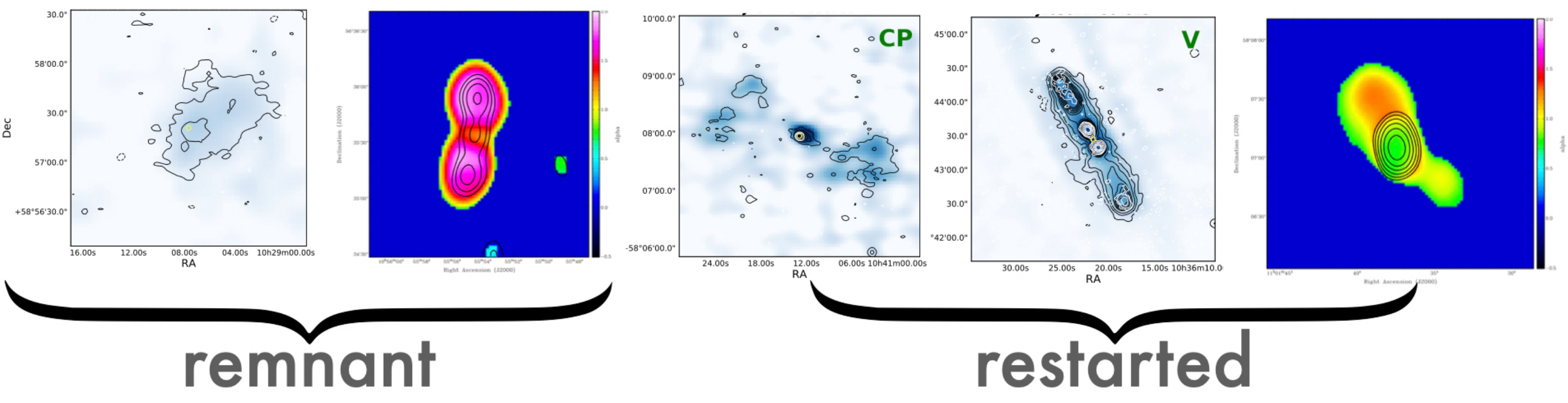}
\caption{Examples of remnant and candidate restarted radio galaxies (from \cite{Jurlin20,Morganti21}). They were selected using different criteria. From left to right: two examples of remnant radio galaxies, the first selected based on the morphology of the emission at low frequencies from the LOFAR image \cite{Jurlin20}, the second based on  USS spectral indices \cite{Morganti21}. The next three panels show  examples of candidate restarted radio galaxies, the first two selected based on  morphology \cite{Jurlin20}, the third  based on spectral index \cite{Morganti21}.
In the total intensity images, the black contours refer to the LOFAR emission, the white contours to FIRST data, and the yellow circle indicates the position of the optical counterpart (taken from \cite{Jurlin21}). The black contours in the spectral index maps represent the 1400~MHz emission from Apertif. As in Figure \ref{fig:NewExamples}, the regions where no contours were drawn indicate regions undetected by Apertif,  corresponding to spectral index lower limits. } 
\label{fig:ExamplesClassification}
\end{figure}

   
\begin{paracol}{2}
\switchcolumn

\section{Exploring the Spatially-Resolved Spectral Indices Distribution of Radio Galaxies in the Lockman Hole Region}
\label{sec:LH}

As the observations of the Apertif surveys proceed, we have now collected data covering 
most of the LOFAR LH field. 
Because of the smaller field-of-view compared to LOFAR and the overlap between different Apertif observations, the images from the beams of 12 Apertif pointings were used to cover the LOFAR LH area and to obtain uniform noise levels over the area. 
The  Apertif images were made using the automatic pipeline, Apercal (Adebahr et al.\ submitted). The present version of this pipeline has a number of limitations (e.g., no direction dependent calibration) and  artefacts are still present around the brightest sources  in some of the images.
However, these artefacts do not affect sources studied here. The quality (and spatial resolution) of the images will improve when a new version of the pipeline (currently being developed) will be available. 

\textls[-15]{The images from all individual Apertif beams were combined---after being convolved to the same angular resolution---into a mosaic image. This was done using the standard script for Apertif data described in Kutkin et al.\ in prep\endnote{https://github.com/akutkin/amosaic}. The mosaic obtained has a resolution of $19 \times 15$ ArcSection. The noise level ranges between 35 and 40 \muJybeam\ in the regions not affected by imaging artefacts. 
The flux scale of the Apertif mosaic was checked against the NVSS, as was previously done for the  image covering a smaller part of the LH area (\cite{Morganti21}). Similarly, a difference of about 10\% in the flux scale was found and corrected for. The flux scale of the LOFAR image was checked and corrected as discussed by \cite{Mandal21}.}

The spectral index image was obtained following the procedure described in \cite{Morganti21}. The LOFAR image was convolved to the same angular resolution as that of the Apertif mosaic, resulting in an image with average rms noise of $\sim$130--160 \muJybeam, increasing toward the edges of the field. The final spectral index image was obtained by combining the spectral index image of the regions detected at both frequencies (obtained by clipping at 5-$\sigma$, i.e.,  0.8 \mJybeam\ for LOFAR, and 0.2 \mJybeam\ for Apertif) with the image of the spectral index limits for regions detected by LOFAR but not by Apertif. This ensures the possibility of revealing regions with (ultra-)steep spectra, something particularly relevant for our characterisation of remnant and restarted sources. 
Figure \ref{fig:RegionSI} shows a small region of the full spectral index image.  This zoom-in clearly illustrates the potential for  resolved spectral index studies as it shows the variety in the spectral index properties of the sources in the field.  
Here, we focus  on the spectral properties of the restarted and remnant radio galaxies that were already selected and included in the sample of \cite{Jurlin20}. The selection of new candidates based only on the spectral properties (as done in \cite{Morganti21} for a small region of the LH) will be done in a future study.
In Figure \ref{fig:NewExamples}, we show some examples of the resolved spectral index distribution in our target sources and below we describe their properties.

We should point out that, { as discussed in more detail in \cite{Morganti21}},  the edges around the sources, where the spectral index appears to become flatter, correspond to those areas where only LOFAR had detected emission, and a fixed upper limit to the Apertif flux \mbox{(5-$\sigma=0.2$ \mJybeam)} was used to calculate the spectral index lower limits. {While the LOFAR flux, as expected, decreases towards the edges of the sources, the Apertif fix flux limit does not: the apparent flattening, therefore, indicates that the lower limit to the spectral index is less stringent at the edges of the sources \citep{Morganti21}}. The flattening is, therefore, not significant.

In the top row of Figure \ref{fig:NewExamples}, three examples of remnant radio galaxies are shown, where USS emission dominates the whole extent of the source, over hundreds of kpc. Interestingly, all three sources are undetected by Apertif and, therefore, the derived spectral indices are lower limits. The actual spectral indices are steeper than what is shown in the figure. The relatively uniform spectral indices throughout the source could be, at least partly, due to the limited angular resolution of the observations, and will need to be further explored with higher spatial resolution observations. However, the observations also indicate that the USS emission  reaches large distances from the core. This suggests that, after the nuclear supply of fresh electrons switches off, emission with increasingly steep spectrum remains visible for a relatively long time (even for sources outside galaxy clusters, such as the three remnant radio sources discussed in this paragraph, see also Section\ \ref{sec:Results}). Ages since the last re-acceleration of the electrons of up to $\sim 300$ Myr have been estimated in \cite{Morganti21}, and are confirmed by the extended USS sources found also by the present work. 

\textls[-15]{The middle row in Figure \ref{fig:NewExamples} shows examples of sources originally selected by \cite{Jurlin20} as candidate restarted radio galaxies, according to the criteria of low CP and low surface brightness of the extended emission. 
The objects in Figure \ref{fig:NewExamples} show that the central region (characterised by a standard spectral index) is surrounded by low surface brightness emission which has a very steep spectral index (steeper than $\alpha \sim$ 1--1.2)  at low frequencies. This suggests the presence of the remnant large-scale emission, possibly originated from a previous phase of activity. 
Thus, the additional information of the spectral index distribution further supports the restarted classification. The extreme spectral indices (in some cases only a limit could be derived so the actual value of $\alpha$ is even steeper) cannot be explained  by only a change in magnetic field between the central region and the USS diffuse large-scale emission. }

\end{paracol}
\begin{paracol}{2}
\switchcolumn


These results confirm the classification scheme of the restarted sources and the robustness of the criteria for selecting these objects. 
The search for steep spectrum  cores (another potential indicator of the presence of restarted emission) will, instead, require higher spatial resolution observations.

The bottom row in Figure \ref{fig:NewExamples} shows other interesting examples illustrating how the resolved spectral index can add additional information about the nature of the sources. These sources were among the candidate remnant radio galaxies originally selected by \cite{Brienza17}, mostly based on low surface brightness and/or low CP from the LOFAR images. However, this classification has been questioned by the follow-up work (using 6000~MHz VLA observations) described in \cite{Jurlin21}, mostly based on the lack of steepening of the spectrum at high frequencies. The complex distribution of the spectral index supports this conclusion, even if regions with USS emission are observed. The structure at 1400~MHz obtained from Apertif, illustrated by the contours in Figure \ref{fig:NewExamples}, clearly appears different from the amorphous morphology of the  confirmed remnant radio galaxies in the top row.  

The images in Figure \ref{fig:NewExamples} show that these sources have cores that are relatively prominent, as seen from their flatter spectral index. Interestingly, the object in the middle of the bottom row shows a region of flatter spectral index  on the E side, possibly tracing a sharp bend of an active jet. In the radio galaxy on the right, the inverted spectrum on the northern edge clearly identifies a background/foreground radio source, likely unrelated to the radio galaxy.  Further multi-frequency observations will be needed to fully understand the nature of these sources. 

Thus, despite being limited to two frequencies, the resolved spectral index images have further confirmed the classification of remnant and restarted radio galaxies and helped in better characterising the properties of these objects, as we will summarise in the next section.

\section{General Findings on Remnant and Restarted Radio Galaxies} 
\label{sec:Results}

Examples of radio galaxies satisfying the various selection criteria for remnant and restarted, are shown in Figure \ref{fig:ExamplesClassification}.
The study of the spectral indices presented in this paper further supports the criteria used. From our studies, we find that the fraction of remnant relative to the comparison sample of active radio galaxies is $\sim 9$\% (\cite{Brienza17,Jurlin20,Jurlin21}). For the candidate restarted radio galaxies, the fraction is more uncertain and ranges between $\sim 7$\% and 15\%  \cite{Jurlin20,Morganti21}. 

\end{paracol}
%

\begin{figure}
\widefigure
\includegraphics[width=18 cm]{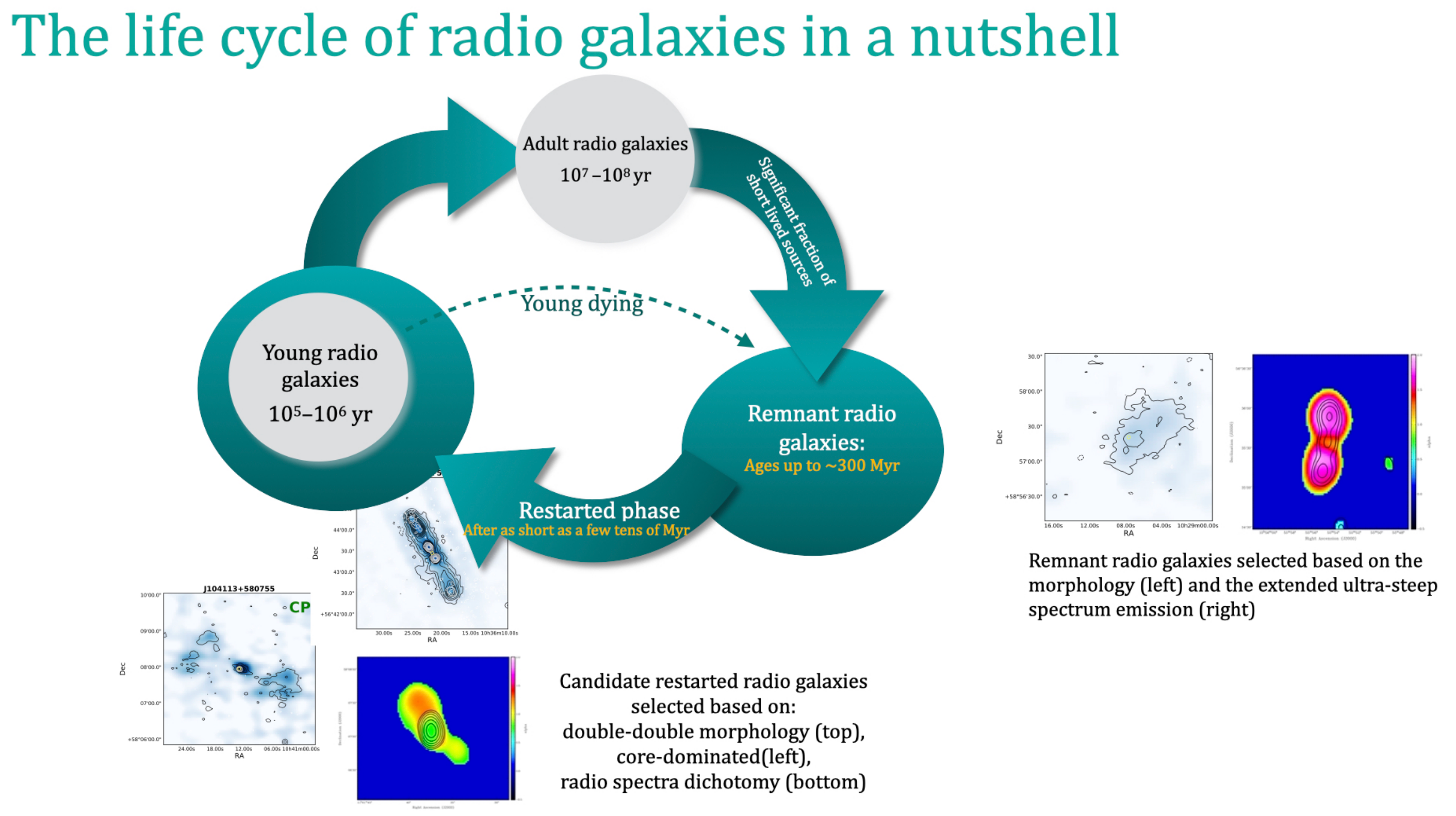}
\caption{Sketch based on the results obtained in the Lockman Hole region by  \cite{Brienza17,Jurlin20,Jurlin21,Morganti21}. See Section\ \ref{sec:Results} for more details. The example of remnant and restarted radio sources are taken from Figure  \ref{fig:ExamplesClassification}. }
\label{fig:LifeCycle}
\end{figure}

\begin{figure}
\widefigure
\includegraphics[width=17 cm]{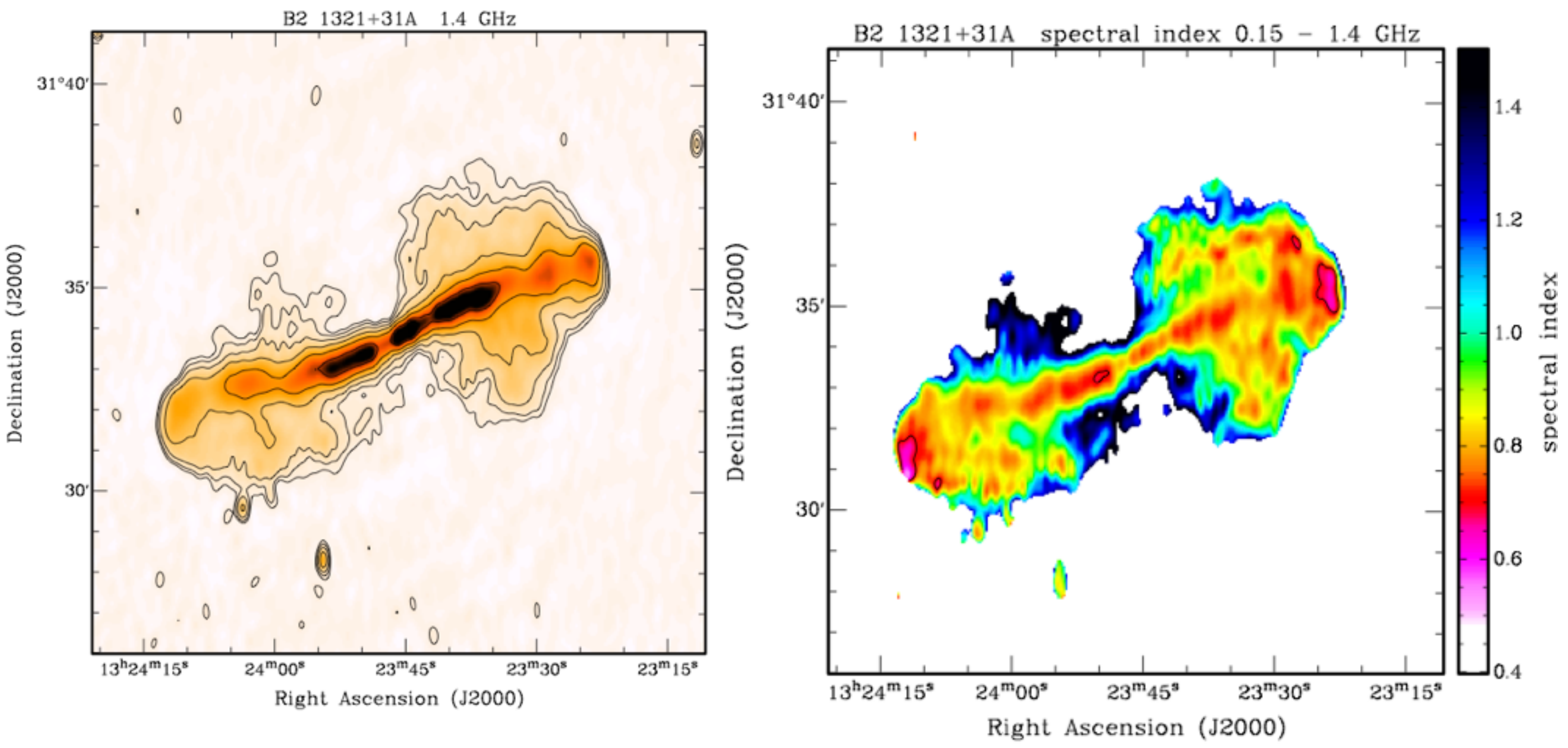}
\caption{B2~1321+31 as seen at 1400~MHz from Apertif (left). On the right, the spectral index image obtained combining Apertif 1400~MHz and LOFAR 150~MHz data. The flatter regions at the end of the jets, reminiscent of hot spots can be seen. 
The steepening toward the central region is also a possible indication of the presence of a backflow, characteristic of FR-II-like sources.}
\label{fig:1321}
\end{figure}

\begin{paracol}{2}
\switchcolumn

The radio luminosities of these objects were found to have median values of log $L_{\rm 150~MHz}/{\rm W\, Hz^{-1}}$ 25.1,  and  25.5  for candidate restarted and remnant radio galaxies, respectively. These values are similar to what is found for sources in the comparison sample (see \cite{Jurlin20} for more details). Thus, the sources in our sample are not representative of the most powerful radio galaxies, and this has likely also implications for the way they evolve. 

We summarise below some of the main results obtained by \cite{Brienza17,Jurlin20,Jurlin21} and from the work presented here. 

\begin{itemize}
\item A number of remnant radio galaxies show USS emission throughout their full extent of many hundreds of kpc, (see, e.g., Figure \ref{fig:NewExamples} and Figure 6 in \cite{Morganti21}). These represent the oldest remnant sources that we detect and have ages  up to  a few hundreds Myr, since the last reacceleration of electrons, derived under the assumption of equipartition conditions between particles and the magnetic field (assuming a typical magnetic field of $\sim 3 \mu G$ \cite{Morganti21}). 

\item Not all remnant sources are characterised by USS emission \cite{Brienza17,Jurlin21}. In a number of cases (including blob 1, see Section\ \ref{sec:SingleObjects}), the steepening of the spectrum only occurs  at high frequencies \cite{Jurlin21}. The  lack of a core (or CP lower than expected for an active radio source of a given radio luminosity), together with the amorphous, low surface-brightness structures, independently confirm the remnant nature of the emission. The spectral properties indicate that there is a variety in the evolutionary stages at which we observe the remnant sources  (e.g., relatively recently switched off vs. extremely old) \cite{Brienza17}.

\item Interestingly, remnant radio galaxies are not  preferentially found in galaxy clusters as earlier suggested (see \cite{Jurlin20,Jurlin21} for more details). Thus, the long timescales during which the remnant  emission can be observed may be related to the fact that these are relatively low-luminosity sources where jet-driven turbulence (and perhaps reacceleration connected to it) plays an important role. This could also result in the relatively uniform distribution of the  spectral index values across the remnant sources. Alternatively, this could be due to the break frequency being below the observed range of frequencies \citep{Murgia11}. The objects in our sample may evolve differently compared to the most powerful radio galaxies modelled by \cite{English19}, typically found to be over-pressured with respect to their environment.  

\item We also confirmed the presence of cores in some of the remnant sources  as it was found for blob 1 (see Section\  \ref{sec:SingleObjects}). Interestingly, some of the USS remnant sources show the presence of a faint radio core (see the discussion in  \cite{Jurlin21}).  This suggests that either the activity never completely stops (i.e., it only strongly dims, perhaps as a result of a change in the SMBH accretion state), or that, at least in some cases, we are seeing the beginning of a restarted phase. These hypotheses will have to be explored using high-resolution images (VLBI, see, e.g., \cite{Parma10}), and by obtaining spectral indices across a broader range of frequencies (e.g., to see whether peaked emission is seen, suggesting the presence of a young radio source). 

\item Related to this, from the resolved spectral indices, we find cases of active emission surrounded by the remnant emission (characterised by the USS). These objects are very interesting  because they suggest that, indeed, the nuclear activity has restarted while the remnant emission from the previous phase of activity is still visible. This indicates that the duty cycle can be relatively fast, estimated around a few tens of Myr. 

\end{itemize}

These results allow us to add some important timescales to the life cycle of radio galaxies, as illustrated in Figure \ref{fig:LifeCycle}, and further discussed below. 
They are also particularly relevant for constraining theoretical modelling of radio source evolution. The models presented in \cite{Brienza17,Godfrey17,Shabala20,Hardcastle18,English19} suggest that the observed fraction of remnant sources can be explained if these structures fade on timescales of at most a few $ \times 10^8$ yr, due to  both radiative losses and dynamical expansion. 
Furthermore, the semi-analytic model described in \cite{Shabala20} shows that a power-law distribution for the ages of the sources ($p(t_{\rm on}) \propto t^{-1}$) is  better at describing the data. The best fit to the data was obtained for a high fraction of short-lived sources (i.e., ${ t_{\rm on} < 100}$ Myr, see \cite{Shabala20} for details).

Thus, as mentioned above, remnant radio galaxies characterised by extended regions of ultra-steep spectrum emission likely represent the oldest tail of the age distribution. The second object from the left in Figure  \ref{fig:ExamplesClassification} shows an example of this. 

On the other hand, the remnant radio galaxies selected based on their morphological properties \cite{Brienza17}, are likely observed soon after  the radio source switched off and are expected to evolve quickly due to dynamic expansion. It is also interesting to note that extreme cases of short-lived radio galaxies (i.e., sources that die before  growing to tens of kpc sizes) have been reported by a number of studies (\cite{Kunert06,Orienti16} and references 
 therein). 

The finding of a duty-cycle that can be relatively fast is further confirmed by the presence of sources in a restarted phase, while the remnant emission is still visible. Timescales of a few tens of Myr, after which a new phase of activity appears to have been derived from their spectral properties (see, e.g., \cite{Brienza18,Brienza20,Morganti21,Kukreti21}) and are confirmed by some of the objects presented in this work. These timescales are consistent with what is derived for double--double radio galaxies, a particular class of restarted sources \cite{Schoenmakers00,Konar13}, as well as with the ages of radio galaxies derived from the work done to build X-ray cavities (e.g., \cite{Vantyghem14}).   
All of these features and timescales are summarised in Figure  \ref{fig:LifeCycle}.

\end{paracol}
\begin{paracol}{2}
\switchcolumn


{Although the results presented here have been mostly obtained thanks to the availability of low frequency surveys, it is important to stress that the multi-wavelength studies, in particular by adding high frequencies, as well as the expansion to high spatial resolution are essential additions for obtaining the complete picture. An other useful addition is the information on polarisation for characterising the conditions of the radio plasma. This information is, at present, limited to a small number of remnant and restarted radio galaxies (see, e.g., \cite{Duchesne19,Adebahr19,Brocksopp11,Orru15}), but it is expected to rapidly expand in the near future, as part of the deliverable of the new large radio surveys.}

\textls[-15]{Going back to the original motivation of this work, it is worth it to connect, at least to first order, the results obtained from the study of the life cycle of radio galaxies with the timescales of the recurrent nuclear activity predicted by numerical simulations of galaxy evolution.}

Many suggestions have been made about the origin of recurrent nuclear activity. They include intrinsic reasons, such as radiation pressure instability within an accretion disk (e.g.,  \cite{Czerny09}),  the cycle of fuelling of the SMBH (e.g.,  \cite{Ciotti10,Gaspari17}), and  external reasons, such as galaxy interactions and mergers. 

The expected timescales of active and non-active phases have been predicted for the cycle of fuelling expected for the Chaotic Cold Accretion model  \cite{Gaspari17}.  According to this model, which has been suggested for jet-mode, inefficient AGN, the intermittency of the activity is connected to the discrete nature of the fuelling clouds, which are formed by condensing out of the hot halo as a result of thermal instabilities. Losing angular momentum via recurrent inelastic collisions, they ‘rain’ down onto the SMBH and feed it.  Thus, it is interesting to see that the predictions of this scenario (see, e.g.,  Figure 15 in \cite{Gaspari17}), show that the balance between heating and cooling creates a loop of about 20--50 Myr between the peaks of activity. This is, to the first order, consistent with what was obtained from our study for the restarting radio AGN.
 
This is promising, although the work done so far is based on limited statistics, limiting also  the possibility to undertake  detailed comparisons with the models. 
Furthermore, refining the derived timescales will require a more realistic description of the ongoing processes.
Nevertheless, these initial results show the potential to dramatically expand this field by making use of the currently available surveys.

\section{Not only Extreme Spectral Indices are Interesting \textellipsis}

As shown by the extensive literature available on the topic, spatially resolved spectral index images can trace the history of the relativistic electrons in a more subtle way than total intensity emission alone. Furthermore, at low frequencies, the spectral signatures can last longer; therefore, providing a `fossil' record of the evolution of the radio source, as described for some examples in Section \ref{sec:SingleObjects}.  

Thus, the properties of the resolved low-frequency spectral index offer a powerful way to trace the history of  radio sources in general and their progenitors, not only limited to remnant and restarted sources. Indeed, the analysis of the spectral properties could also be relevant for better tracing the Fanaroff--Riley classification of radio galaxies, the details of which have been recently questioned by \cite{Capetti17,Mingo19}.  Using the LOFAR survey, \cite{Mingo19} have shown the presence of a population of low-luminosity radio galaxies with FR-II morphology. A possible scenario to explain the origin of this population is that these low-luminosity FR-II objects are older sources, which have begun to fade from their peak radio luminosity (e.g.,~\cite{Shabala08}): the spectral properties can be useful for exploring this scenario.

A possible test case is the radio galaxy B2~1321+35, whose low frequency spectral index map we present here for the first time.
For study of the spectral properties of this object, we used the images  obtained from the  Apertif and LOFAR surveys, as described in Section\ \ref{sec:surveys}.  Thus, this work illustrates the good quality of the data, now mostly public, and that detailed spectral index studies can be routinely done with these data.

The radio morphology of B2~1321+35 observed by Apertif at 1400~MHz and the spectral index between 150--1400 MHz are shown in Figure \ref{fig:1321}. 
This object has been discussed in a number of studies \cite{Klein95,Parma99}.
Interestingly, it was found by \cite{Klein95} to have a spectral index steepening (between 49 and 2.8 cm) in the direction perpendicular to the major axis. The possibility of this indicating  `leaking' across the boundary of the radio jets was suggested by \cite{Fanti82,Klein95}.
The spectral index distribution at low frequencies obtained by combining the 150 MHz data from LOFAR and the 1400 MHz image of Apertif, indeed shows particularly steep spectra  away  from the jet axis. However, the interesting feature from the low-frequency spectral index is the possible presence of flatter regions at  the end of the radio jets, reminiscent of what usually seen in FR-II sources. These regions (in pink in \mbox{Figure \ref{fig:1321}}) have average spectral index values $\alpha^{150}_{1400} \sim 0.65$, flatter than any other structure in the source. 
Backflow of plasma (i.e., stream-like structures) is also seen emerging from these regions to the southeast  and northwest of the lobes.

The spectral properties are interesting and puzzling because the source does not show hot spots in total intensity (see the left panel in Figure  \ref{fig:1321}) and following the classical definition of surface brightness ($\geq$0.6 mJy arcsec$^{-2}$) and surface brightness contrast with respect to the entire lobe ($\geq$4, \cite{deRuiter90}), it clearly shows a (lobed) FR-I morphology.   The radio luminosity at 1400~MHz of $\sim 10^{24}$ W Hz$^{-1}$, is also typical of this type of sources.

Lobed FR-I radio galaxies with similar properties have been seen before, in particular in \cite{Laing11}, where in some of the sources, `cap' regions were identified at the end of the jet, sometimes  associated with the region of flatter spectral indices.
This is clearly different from what happens in the `tailed' FR-I radio galaxies (like 3C~31), where the spectral index steepens at larger distances from the radio core.
Our results for B2~1321+31 show that detailed spectral index distribution can be now routinely  derived for all of the radio galaxies in the area covered by LOFAR and Apertif. 

\end{paracol}

\begin{paracol}{2}
\switchcolumn

The flattening seen in the region at the end of the jet for a (lobed) FR-I of low-power object, such as B2~1321+31, could perhaps be connected to the population of low-luminosity radio galaxies with FR-II morphology, highlighted by \cite{Capetti17,Mingo19} as described above.  When adding the spectral index information, we could speculate that B2~1321+31 may represent an example of such low-luminosity FR-II that has already aged and is found in an advanced stage of evolution. 
In B2~1321+31, the fading of the FR-II structure has resulted in a lower luminosity, but also in fading of the hot spots, which now can only be recognised in the spectral properties at low frequencies.
Further support to this interpretation is the very weak VLBI core (peak flux 3.7~mJy; \cite{Giovannini05}) which suggests a  core prominence lower than what typically observed in FR-I.

With the LOFAR and Apertif surveys, this kind of studies can now be routinely done over large areas and for many different types of sources. The low-frequency spectral index can be investigated  over the entire overlapping area of the LOFAR and Apertif observations at an unprecedented depth for such a large sky area and  with good angular resolution. This offers the opportunity to investigate in more detail the evolutionary stage of a source in connection with the morphological classification. This would allow to investigate whether the spectral properties found in B2~1321+31 are  common in extended radio galaxies and, if so, for which population.

\section{Conclusions}

In this paper, we summarised some of the recent results on the study of the life cycle of radio galaxies, obtained by focusing on the elusive phases of remnant and restarted radio sources. The results presented here were obtained by using LOFAR images at 150~MHz---for the selection based on the morphology---in combination with Apertif images at 1400~MHz to derive spatially resolved, spectral information at low frequencies. 
Our results show that these sources have a larger variety of properties compared to what was previously found. This variety is likely the result of differences in their  evolutionary stage and their progenitors.

Many radio sources  with extended USS emissions were found. This can be used for understanding, in more detail, the evolution of the radio emission after the nuclear activity stops. Interestingly, we are also finding objects with USS emission, surrounding a relatively bright radio core.  We identify these as restarted radio sources. They could indicate a relatively fast duty-cycle for switching on and off. If this is confirmed, it would have important implications for understanding the role of feedback from radio AGN.

The resolved spectral index images can also provide important information on the evolution of  radio sources, especially now that recent results have questioned the assumption of a sharp  separation between FR-I and FR-II radio galaxies.  The large and deep low- and mid-frequency surveys that are becoming available make it possible to use the spectral information for a better understanding of this classification. 

The results presented here are limited to high-band antenna (HBA)  LOFAR observations at 150~MHz. However, data from low-band array (LBA) at even lower frequency \mbox{($\sim$60 MHz)} are now becoming available \cite{deGasperin21}, which is highly relevant for the topics discussed here. These data have a very similar spatial resolution and comparable depth (for steep spectrum sources) as the Apertif survey, making an ideal combination for extending the spectral index studies to the range 60--1400~MHz.

\vspace{6pt} 

\authorcontributions{Conceptualization and formal analysis, R.M., N.J., M.B., T.O.; software, A.K., J.Z.; data curation, E.A.K.A., E.O., A.K., H.D., K.H.; writing---original draft preparation, R.M.; writing---review and editing, T.O., M.B., I.P., J.M.v.d.H., A.S. All authors have read and agreed to the published version of the manuscript.}

\funding{The research, leading to these results, received funding from the European Research Council under the European Union's Seventh Framework Programme (FP/2007-2013)/ERC Advanced Grant RADIOLIFE-320745. M.B. acknowledges support from the ERC-Stg ``DRANOEL", no. 714245, from the ERC-Stg ``MAGCOW", no. 714196.
E.A.K.A. is supported by the WISE research programme, which is financed by the Netherlands Organization for Scientific Research (NWO).
J.M.v.d.H. acknowledges funding from the Europeaní Research Council under the European Union’s Seventh Framework Programme (FP/2007-2013)/ERC grant agreement no. 291531 (‘HIStoryNU’).
}

\institutionalreview{Not applicable}

\informedconsent{Not applicable
}


\dataavailability{LOFAR 150 MHz image of the LH has been release in combination with the paper of \cite{Tasse21}.
The Apertif images from the first data release are available (since 10 Nov 2020) at \url{http://hdl.handle.net/21.12136/B014022C-978B-40F6-96C6-1A3B1F4A3DB0}.} 

\acknowledgments{This work makes use of data from the Apertif system installed at the Westerbork Synthesis Radio Telescope owned by ASTRON. ASTRON, the Netherlands Institute for Radio Astronomy, is an institute of the Dutch Science Organisation (De Nederlandse Organisatie voor Wetenschappelijk Onderzoek, NWO).
LOFAR, the Low Frequency Array designed and constructed by ASTRON, has facilities in several countries, which are owned by various parties (each with their own funding sources), and that are collectively operated by the International LOFAR Telescope (ILT) foundation, under a joint scientific policy. }

\conflictsofinterest{The authors declare no conflict of interest.} 

\abbreviations{Abbreviations}{The following abbreviations are used in this manuscript:\\

\noindent 
\begin{tabular}{@{}ll}
AGN& active galactic nucleus\\
Apertif & APERture Tile In Focus \\
CP&core prominence\\
FIRST&faint images of the radio sky at twenty-centimeters\\
FR-II& Fanaroff--Riley class II\\
GMRT & giant metrewave radio telescope \\
HBA& high-band array\\
LBA& low-band array\\
LH& Lockman Hole
\end{tabular}
 
\noindent\begin{tabular}{@{}ll}
LoLSS & LOFAR LBA Sky Survey\\
LOFAR & LOw Frequency ARray\\
LoTSS& LOFAR Two-Metre Sky Survey\\
NVSS& The NRAO VLA Sky Survey\\
PAF & phased-array feed \\
SMBH & super massive black hole\\
SED&spectral energy distribution\\
SPC&spectral index curvature\\
USS&ultra steep spectra\\
VLA& very large array\\
VLBI & very long baseline interferometry\\
WSRT & Westerbork synthesis radio telescope
\end{tabular}}

%




\end{paracol}
\printendnotes[custom]
\reftitle{References}

\end{document}